\documentclass[journal]{IEEEtran}
\pdfoutput=1
\ifCLASSINFOpdf
   \usepackage[pdftex]{graphicx}
   \DeclareGraphicsExtensions{.pdf,.jpeg,.png,.eps}
\else

   \usepackage[dvips]{graphicx}
\fi
\usepackage[cmex10]{amsmath}

\usepackage{array}
\usepackage{mdwmath}
\usepackage{mdwtab}
\usepackage{eqparbox}
\usepackage{adjustbox}
\usepackage{booktabs}
\usepackage{color}

\usepackage[tight,footnotesize]{subfigure}
\usepackage{multirow}
\usepackage{cite}
\usepackage{stfloats}

\begin{document}
\title{Design of a $\beta$-Ga$_2$O$_3$ Schottky Barrier Diode With p-type III-Nitride Guard Ring for Enhanced Breakdown}
 
\author{Saurav~Roy,
        ~Arkka Bhattacharyya,
        ~and Sriram Krishnamoorthy      

 \vspace{-0.42cm}
\thanks{Saurav Roy, Arkka Bhattacharyya, and Sriram Krishnamoorthy are with the Department of Electrical and Computer Engineering, The University of Utah, Salt Lake City, UT, 84112, United States of America (e-mail: u1268405@utah.edu; a.bhattacharyya@utah.edu; sriram.krishnamoorthy@utah.edu).}}


\maketitle
\begin{abstract}
This work presents the electrostatic analysis of a novel Ga$_2$O$_3$ vertical Schottky diode with three different guard ring configurations to reduce the peak electric field at the metal edges. Highly doped p-type GaN, p-type nonpolar AlGaN and polarization doped graded p-AlGaN is simulated and analyzed as the guard ring material, which forms a heterojunction with the Ga$_2$O$_3$ drift layer. Guard ring with non-polar graded p-AlGaN with band gap larger than Ga$_2$O$_3$ is found to show best performance in terms of screening the electric field at the metal edges. The proposed guard ring configuration is also compared with a reported Ga$_2$O$_3$ Schottky diode with no guard ring and a structure with high resistive Nitrogen doped guard ring. The optimized design is predicted to have breakdown voltage as high as 5.3 kV and a specific on resistance of 3.55 m$\Omega$-cm$^2$ which leads to an excellent power figure of merit of 7.91 GW/cm$^2$.   
\end{abstract}

\IEEEpeerreviewmaketitle

\section{Introduction}
\label{sec1}
\IEEEPARstart{G}{allium Oxide} (Ga$_2$O$_3$) has a huge potential for power device applications due to its high breakdown field. $\beta$-Ga$_2$O$_3$ has a band gap (4.6 eV) larger than GaN and SiC, with an estimated critical breakdown field as high as 8 MV/cm. Due to the large critical electric field, the Baliga Figure of Merit (BFOM) relevant to power switching could be 2000–3400 times that of Si, which is several times larger than that of SiC or GaN. Low doped drift layers in conjunction with large band gap materials can enable very high breakdown voltage. Various power devices using $\beta$-Ga$_2$O$_3$ have been demonstrated recently with high breakdown voltage in the vertical geometry. \cite{higashiwaki2012,sasaki2012device,sasaki2013hbox,higashiwaki2015ga,konishi20171,yang2017highb,sasaki2017first,yang2017high,yang20182300v,li2018new,li20182,wang2019high}.

\begin{figure}[t]
\centering
\includegraphics[width=3.5in,height=7cm]{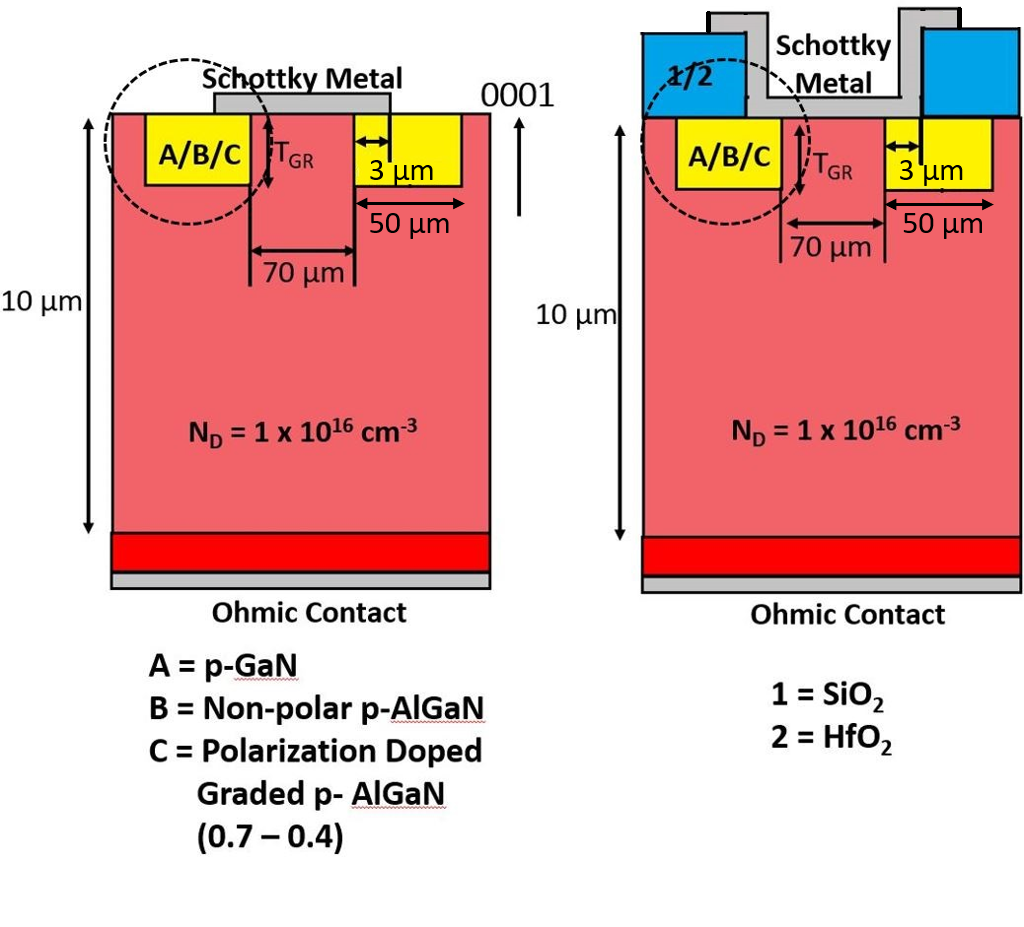}
\caption{Schematic of Ga$_2$O$_3$ SBD with guard rings (GR) where (A) Non-polar p-GaN GR (B) Non-Polar p-AlGaN GR (C) Polarization doped graded p-AlGaN GR with (1) SiO$_2$ (2) HfO$_2$ as field plate oxide.}
\label{fig1}
\vspace{-0.5cm}
\end{figure}

Several field management techniques have been explored for Schottky diodes over the years that includes edge terminations, superjunctions etc. Guard ring is one such edge termination technique where the anode metal edge is surrounded by a doped region with opposite polarity to that of the drift region to screen the high electric field generated at the metal edge. Lin et. al. \cite{lin2019} recently demonstrated a Schottky barrier diode (SBD) with nitrogen ion implanted guard ring (GR) with a maximum breakdown voltage of 1.43 kV. Zhou et. al. \cite{zhou2019} also demonstrated a Ga$_2$O$_3$ SBD using Mg ion implanted guard ring with a breakdown voltage of 1.65 kV. A similar design with Argon implanted edge termination have also been reported by Gao et. al. \cite{gao2019high}. Although these devices can achieve a breakdown improvement compared to the case with no guard rings, the lack of electric field screening due to the absence of carriers in the guard ring limits the breakdown voltage. A high resistive guard ring, as demonstrated in the previous devices can spread the depletion region and can reduce the field crowding at the metal edges. However, a guard ring with mobile holes can be very effective in screening the electric field at the metal edges due to the presence of a quasi neutral region and can dramatically shift the high field region from the metal edge to deep inside the device, thereby eliminating the effect of surface states which causes premature breakdown. Because of the difficulty in having  hole concentration in $\beta$-Ga$_2$O$_3$ due to the absence of a shallow acceptor and hole self trapping, p-doped III-Nitrides would be a viable option to get a reasonably high hole concentration.  The idea of heterostructure guard rings have been proposed on Silicon Carbide substrate previously \cite{zhang2015semiconductor}. \color{black}  Muhammed et. al. \cite{muhammed2016} have reported the growth of c-plane n-GaN epilayer on ($\Bar{2}$ 0 1) $\beta$-Ga$_2$O$_3$ substrate using MOCVD. Vertical blue LEDs have also been demonstrated on ($\Bar{2}$ 0 1) $\beta$-Ga$_2$O$_3$ substrates \cite{muhammed2017}.  Shimamura et. al. \cite{shimamura2004} reported growth of c-plane GaN on (1 0 0) oriented $\beta$-Ga$_2$O$_3$ substrate using MOCVD. On the other hand Cao et. al. \cite{cao2017} reported the growth of non-polar a-plane GaN on (0 1 0) oriented $\beta$-Ga$_2$O$_3$ substrate by MOCVD. All these reports confirm the viability of growing electronic grade polar and non-polar GaN on $\beta$-Ga$_2$O$_3$ substrates depending on the choice of orientation of the substrate.

\begin{table} []
\caption{Material parameters}
\begin{center}
\begin{tabular}{ |c|c|c|c| }
\hline
 Material & Ga$_2$O$_3$ & GaN & AlN \\ 
 \hline \hline
 Bandgap (eV) & 4.85 \cite{pearton2018review} & 3.3 \cite{vurgaftman2001band} & 6.2 \cite{vurgaftman2001band}  \\  
 \hline
 Electron affinity (eV) & 3.9 \cite{pearton2018review} & 3.9 \cite{vurgaftman2001band} & 0.6 \cite{vurgaftman2001band}\\
 \hline
 Relative permittivity & 10 \cite{pearton2018review} & 8.9 \cite{Levinshtein2001PropertiesOA} & 8.5 \cite{Levinshtein2001PropertiesOA}  \\
 \hline
 Electron Effective mass & 0.28m$_0$  & 0.22m$_0$  & 0.4m$_0$ \\
 \hline
 Hole Effective mass & - & 1m$_0$  & 4m$_0$ \\
 \hline
 Critical Electric Field (MV/cm) & 8 \cite{pearton2018review} & 3.3 \cite{Levinshtein2001PropertiesOA} & 15 \cite{Levinshtein2001PropertiesOA} \\
 \hline
 Sn Activation energy (eV) & 0.13 \cite{pearton2018review} & - & -\\
 \hline
 Mg Activation energy (eV) & - & 0.2 \cite{Levinshtein2001PropertiesOA} & 0.6 \cite{Levinshtein2001PropertiesOA}\\
 \hline
 Spontaneous Polarization (C/m$^2$) & - & -0.034  & -0.09  \\
 \hline
\end{tabular}
\label{t1}
\end{center}
\end{table}
\color{black}

\begin{figure}[t]
\centering
\subfigure[]{
\includegraphics[width=3.2in, height=1.4in]{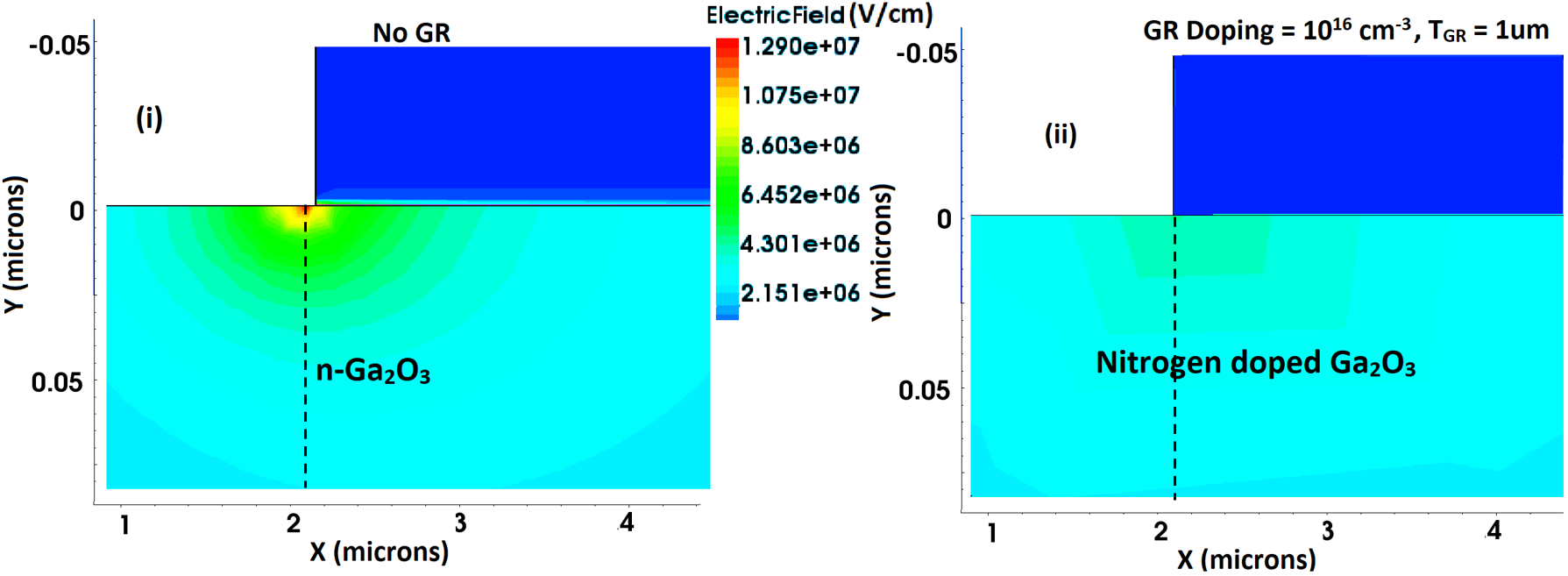}}\hspace{-0.2cm}
\subfigure[]{
\includegraphics[width=3.1in, height=2.5in]{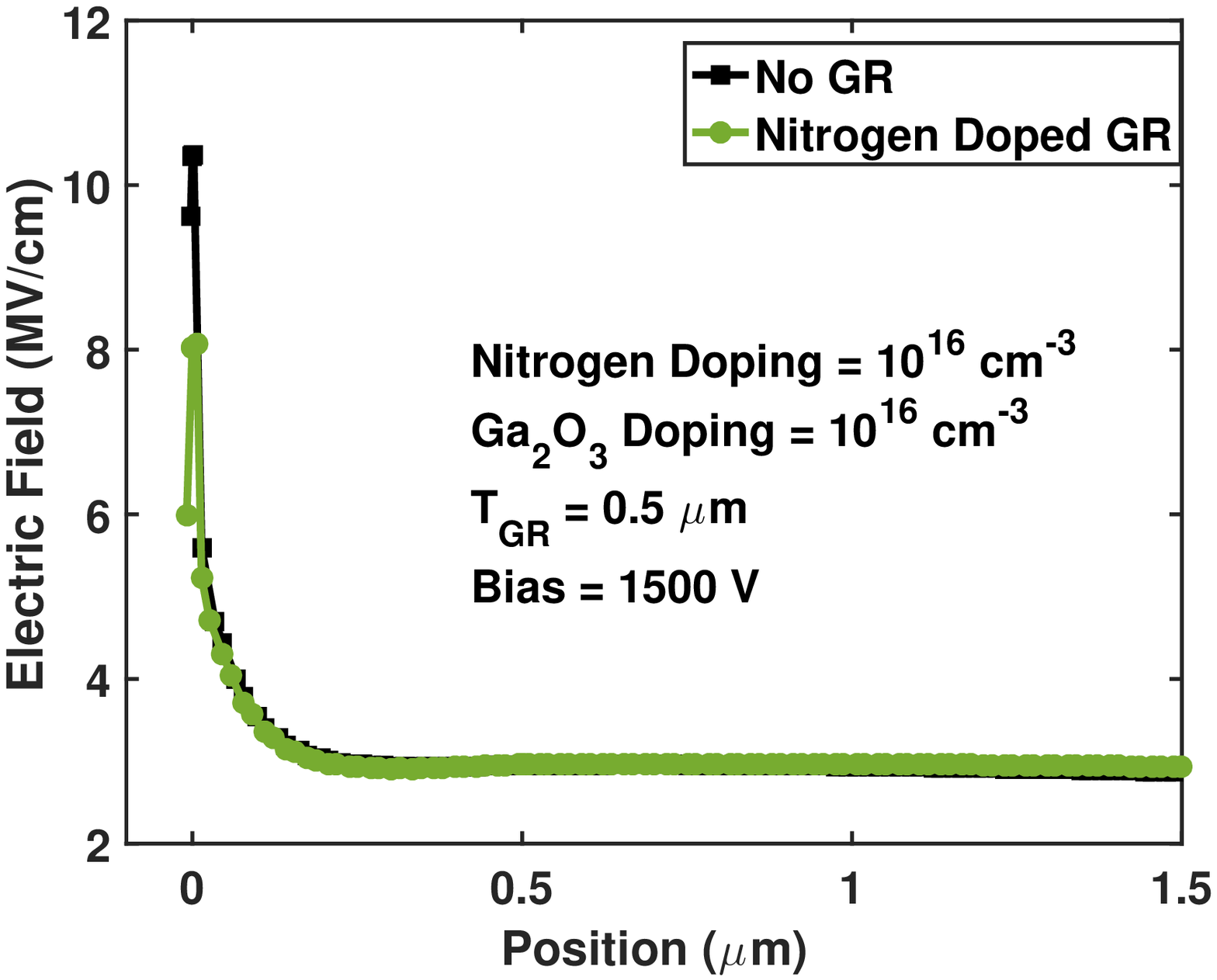}}\hspace{-0.2cm}
\caption{\footnotesize (a) Electric field distribution in SBD for (i) no GR (ii) Nitrogen doped GR at bias voltage of 1500 V (b) Electric field vs position for N-doped GR and no GR at 1500 V along the cutlines shown in the contour plots. }
\label{fig2}
\vspace{-0.2cm}
\end{figure}

\begin{figure}[t]
\centering
\subfigure[]{
\includegraphics[width=1.71in, height=1.35in]{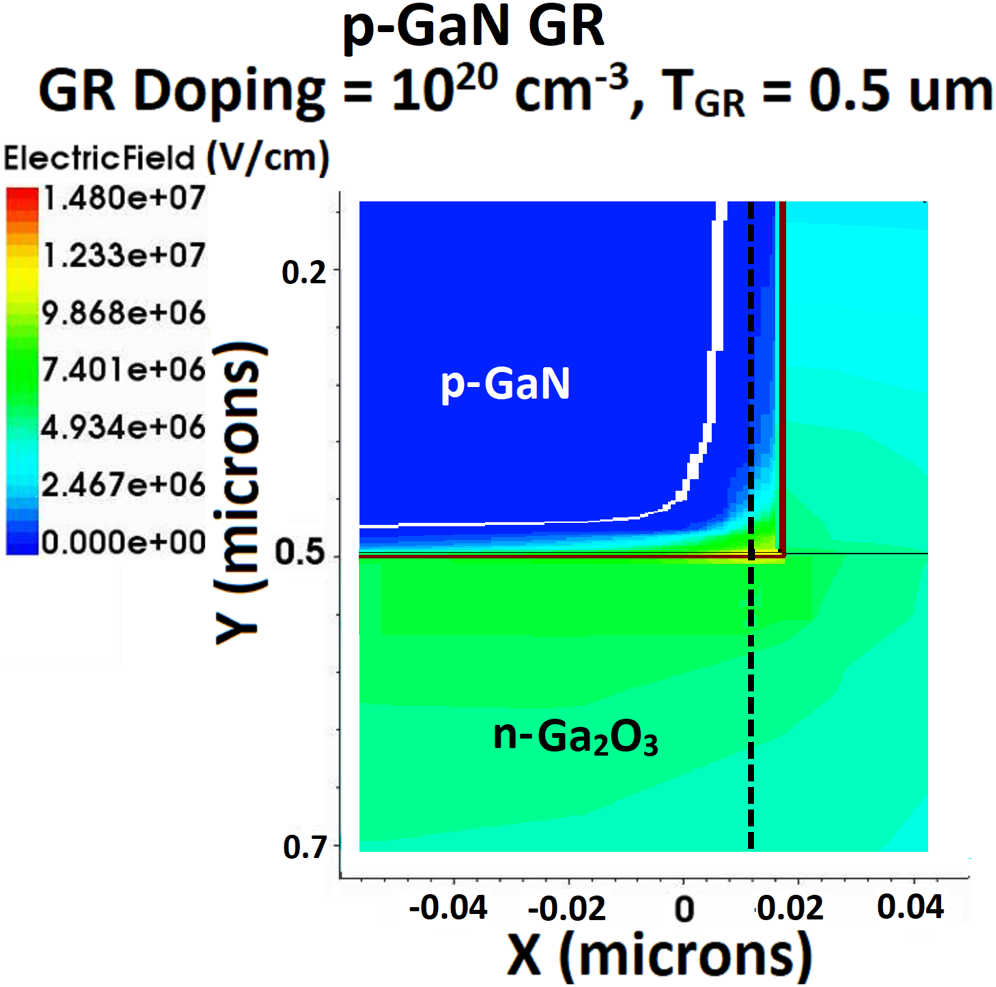}}\hspace{-0.2cm}
\subfigure[]{
\includegraphics[width=1.71in, height=1.35in]{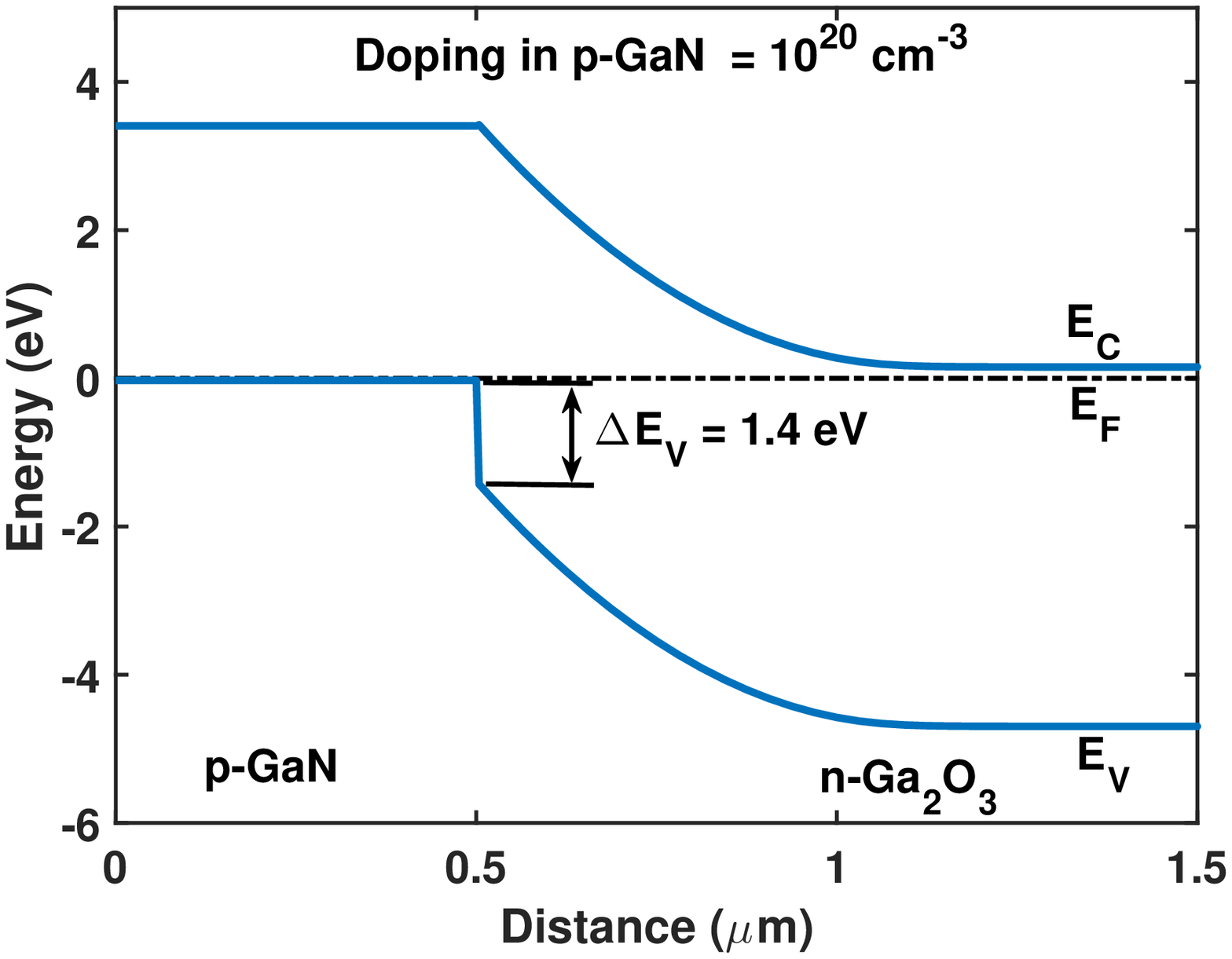}}
\subfigure[]{
\includegraphics[width=1.71in, height=1.35in]{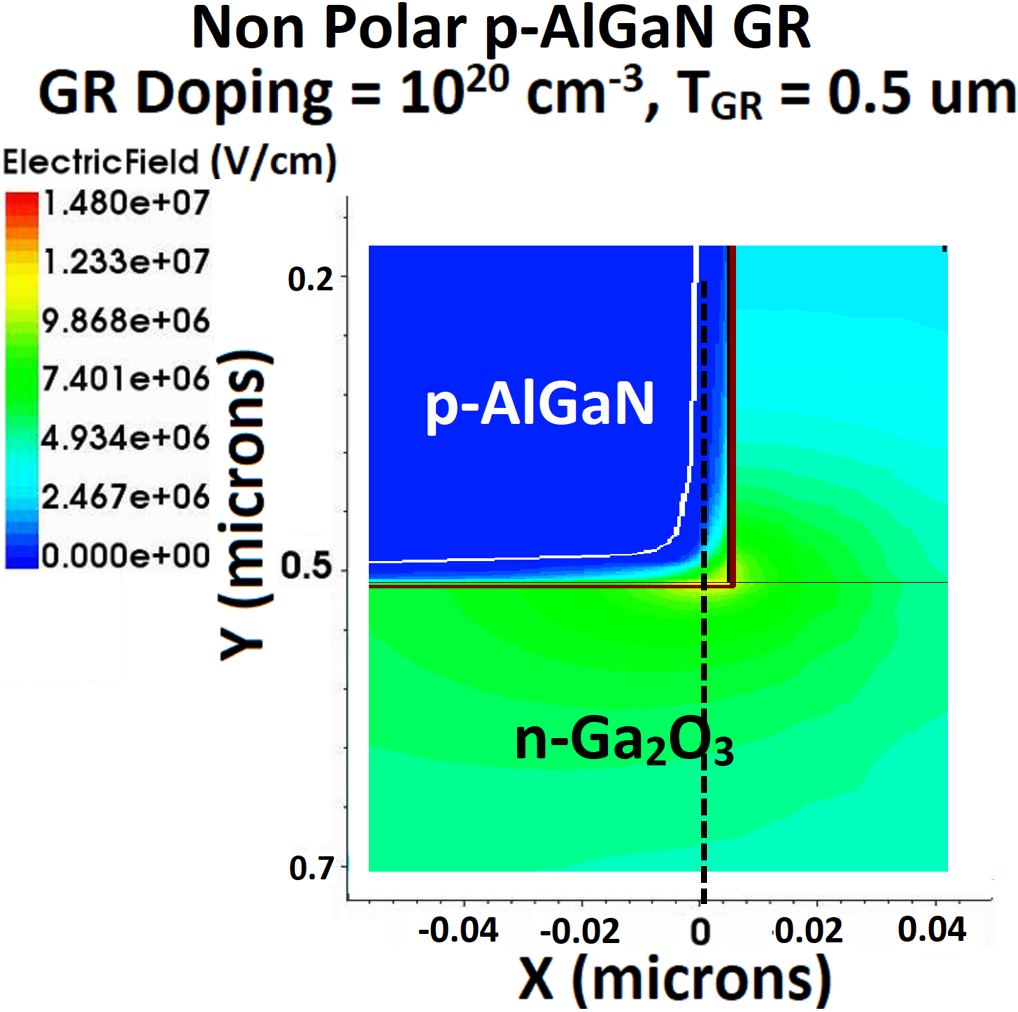}}\hspace{-0.2cm}
\subfigure[]{
\includegraphics[width=1.71in, height=1.35in]{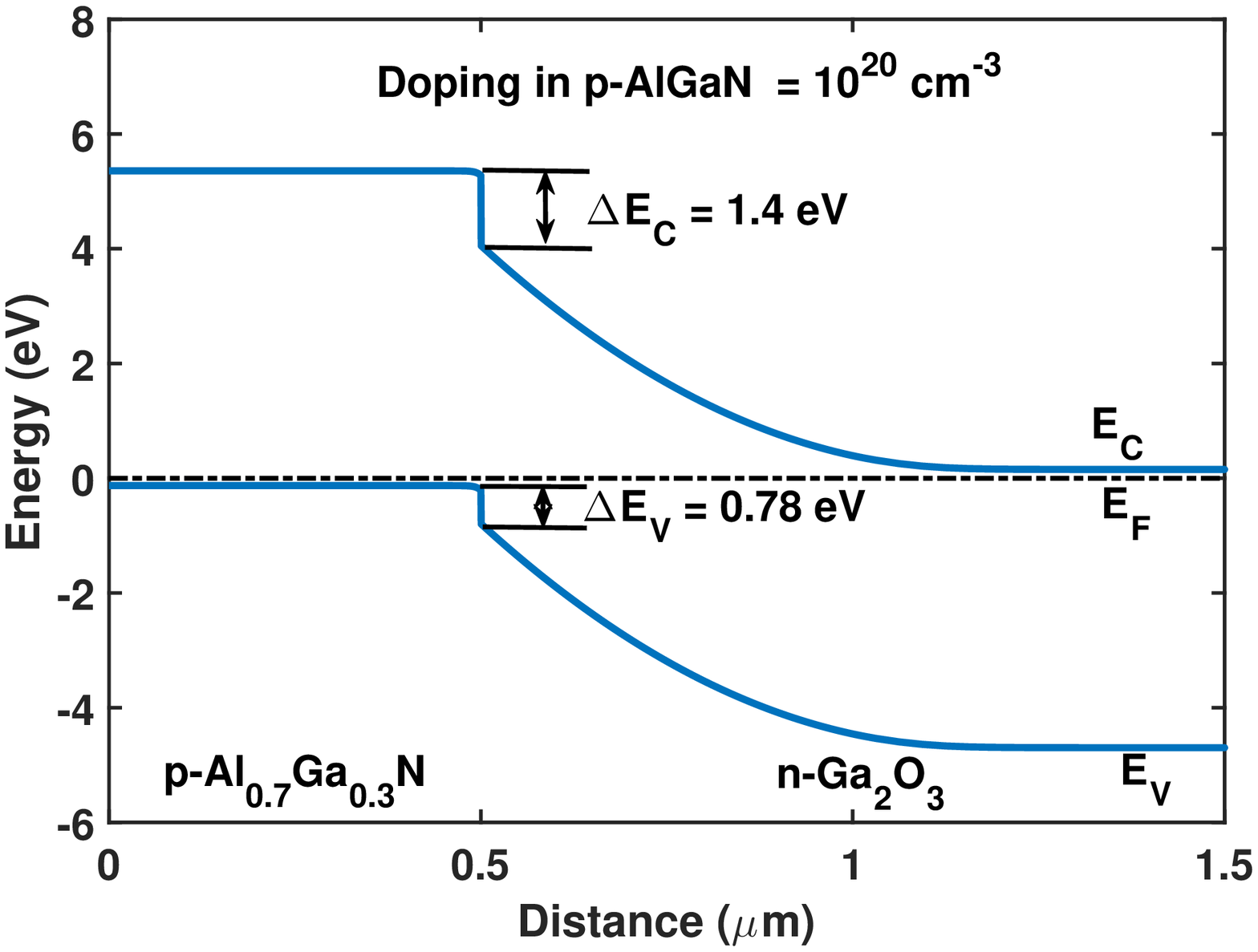}}\hspace{-0.2cm}
\subfigure[]{
\includegraphics[width=1.71in, height=1.35in]{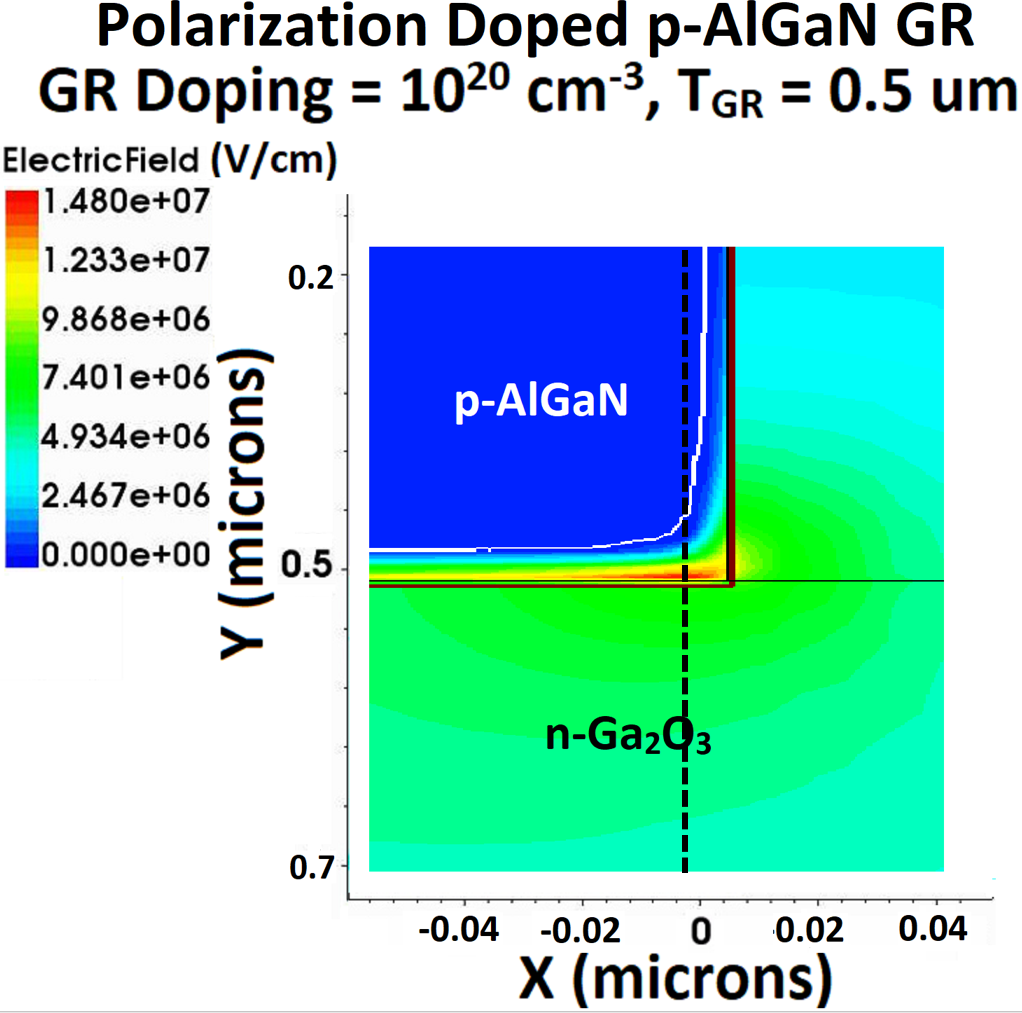}}\hspace{-0.2cm}
\subfigure[]{
\includegraphics[width=1.71in, height=1.35in]{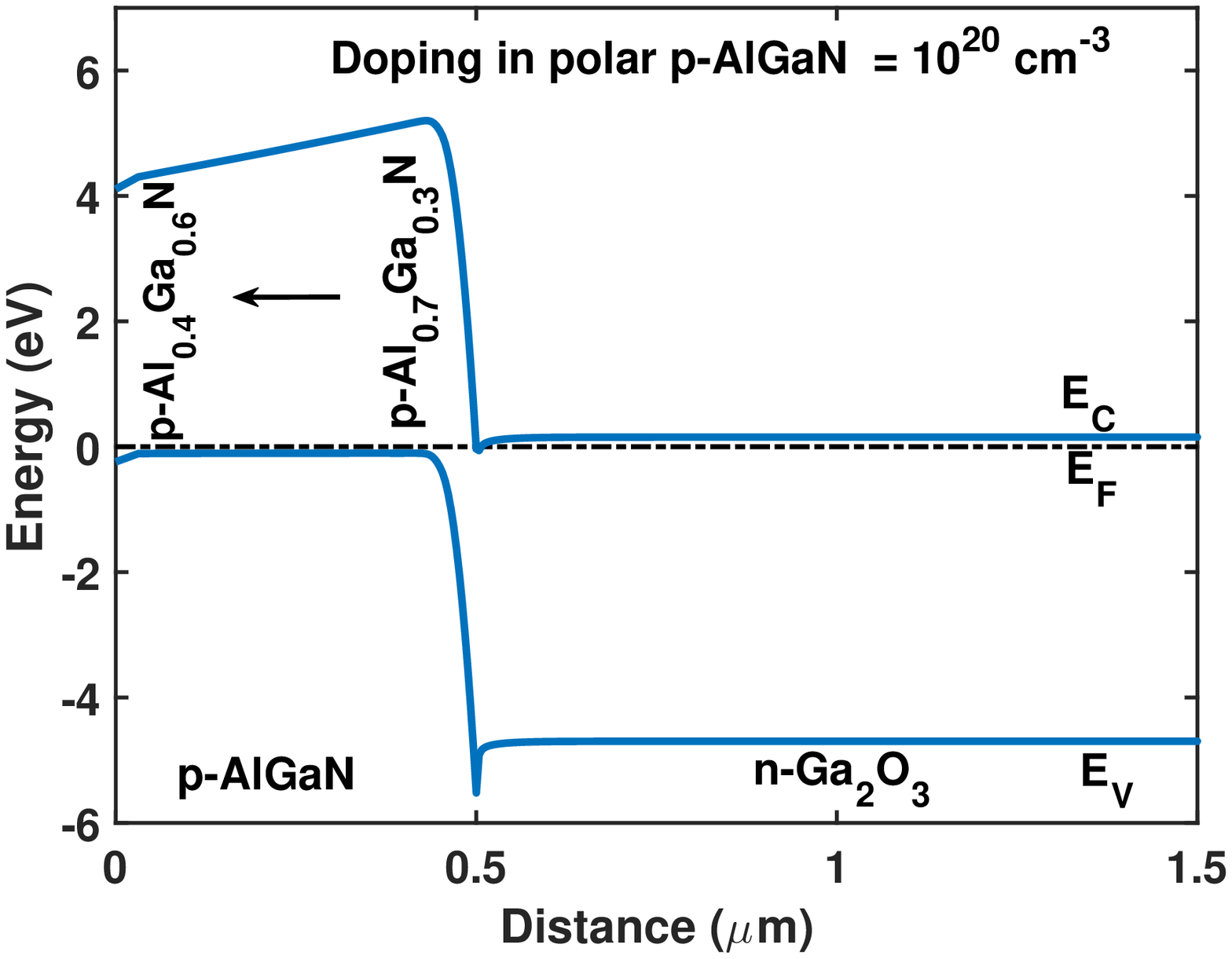}}\hspace{-0.2cm}
\\
\subfigure[]{
\includegraphics[width=2.6in, height=2.1in]{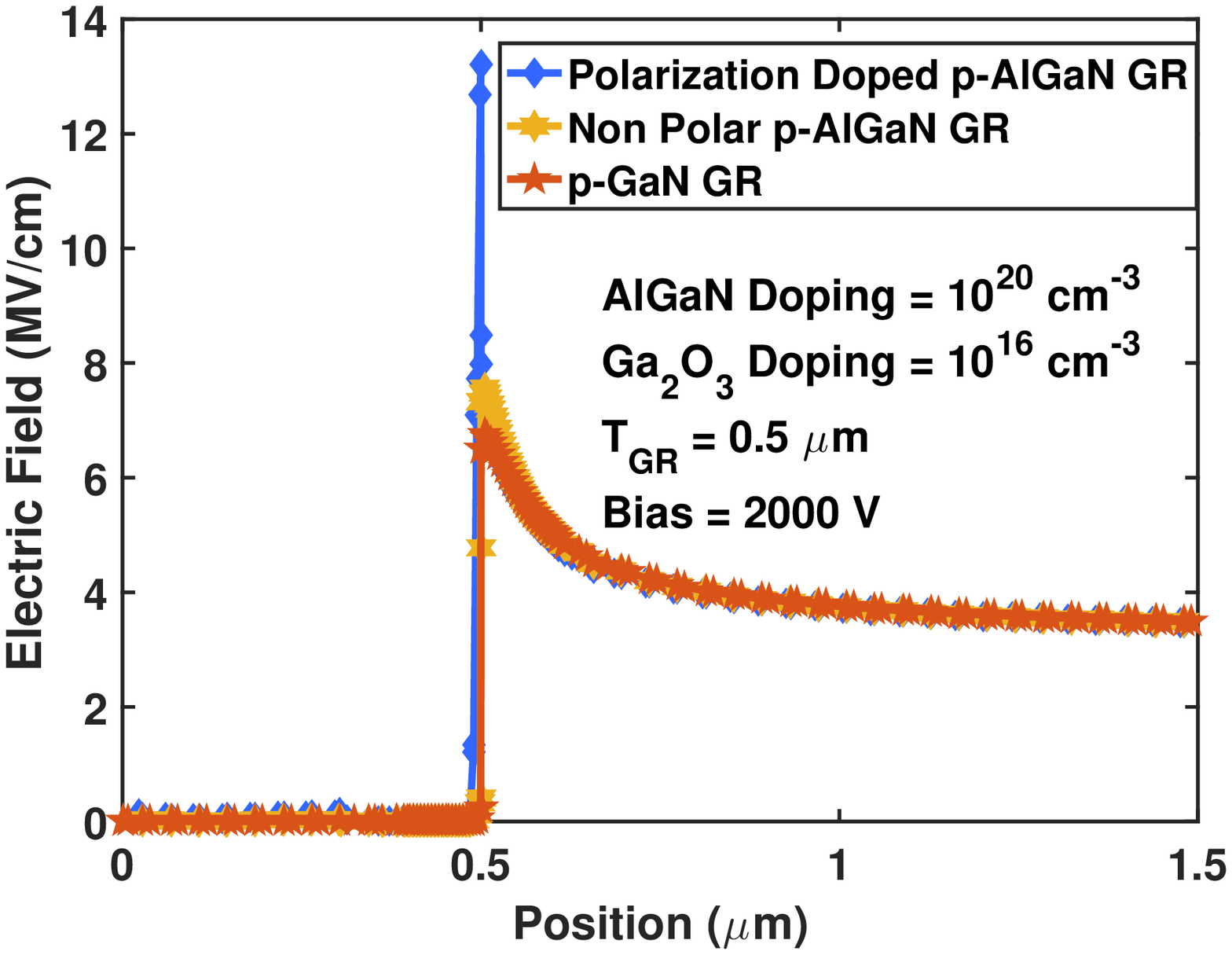}}
\caption{\footnotesize Electric field distribution and equilibrium energy band diagram for SBD with (a) (b) non-polar p-GaN GR (c) (d) non-polar p-AlGaN GR (e) (f) polarization doped p-AlGaN GR respectively at bias voltage of 2000 V and GR doping of $10^{20}$ $cm^{-3}$ (g) Electric field vs position in SBD for the three different GR at 2000 V and at GR doping of $10^{20}$ $cm^{-3}$ along the cutlines shown in the contour plots.}
\label{fig3}
\end{figure}


In this paper, we propose and design a Ga$_2$O$_3$ SBD with a p-doped III-Nitride guard ring using electric field simulations. We have explored three guard ring configurations including (i) p-Gallium Nitride (p-GaN) GR, (ii) Non-polar graded p-Aluminum Gallium Nitride (p-AlGaN) GR and a (ii) polar graded p-AlGaN GR. In this work we perform detailed electrostatic simulations to capture and manage high in III-nitride/$\beta$-Ga$_2$O$_3$ heterostructures. The design is optimized to extract the optimum device parameters which efficiently reduces the electric field. We have also explored the additional use of field plate in the aforementioned optimum design to further minimize the peak electric field in the device structure. 


\vspace{-0.2cm}
\section{Simulation Methodology}
\label{sec1a}
The schottky barrier diode device structure (Fig. \ref{fig1}) with varied guard ring thickness of T$_{GR}$, a 10 $\mu m$ thick drift layer (N$_D$ = $10^{16}$ cm$^{-3}$), Ni/Au Schottky metal with a barrier height $\Phi_B$ of 1.4 eV \cite{arkka1}, and a Schottky metal guard ring overlap of 3 $\mu$m is simulated using Sentaurus \cite{sentaurus} 2D TCAD device simulator. Width of the guard ring is considered to be 50 $\mu$m. In the simulation, adequate numerical convergence was reached by an optimized meshing, with subnanometer grid spacing for the key electrical layers and their interfaces and larger spacings for drift region. Spontaneous polarization model is used to include polarization effect in the case of graded polar p-AlGaN.The sponataneous polarization values of -0.034 and -0.09 C/m$^2$ is considered for GaN and AlN \cite{Levinshtein2001PropertiesOA}. For p-type doping in GaN and AlGaN, incomplete ionization model is also used to reflect the accurate hole concentration. In order to capture the accurate results for high doping case, Fermi-Dirac model is included for device operating biases. The device simulation setup uses well calibrated mobility model and thermodynamic transport model to match the recent experimental results \cite{lin2019}. The device breakdown voltage can be extracted from E-field simulation when the peak E-field reaches the GaN (3.3 MV/cm) or Ga$_2$O$_3$ (8 MV/cm) critical E-field. Band offsets were determined using electron affinity rule. The GaN/Ga$_2$O$_3$ band offset estimated using electron affinity rule matches well with the experimentally determined band offsets \cite{wei2012valence}. The ionization integrals for avalanche breakdown were not evaluated in order to avoid excessive computation time. Furthermore, accurate ionization rate parameters are currently unknown for Ga$_2$O$_3$. Hence it should be noted that the breakdown is not directly calculated, but can be estimated based on the generated electric field distributions \cite{xiao2019,zhang2013,xia2019design}. It should also be noted that in real devices field crowding can also occur in the device corners. However those fields are always lower than the field crowding at the electrode edges which is the primary cause of device breakdown. Hence our comparison of device breakdown for the various configurations based on 2-D simulation is considerably reliable and was also demonstrated by other works \cite{xiao2019} .All the material parameters for $\beta$- Ga$_2$O$_3$, GaN and AlN assumed in the device simulation are presented in Table \ref{t1}. All the other material parameters for intermediate Al compositions in Al$_x$Ga$_{1-x}$N have been calculated based on GaN and AlN parameters using Vegard's law.

\vspace{-0.2cm}
\section{Results and Discussions}
\label{sec3}
The SBD with nitrogen doped GR and the one with no GR is simulated for comparison and the electric field profiles are shown in Fig. \ref{fig2}(a) (i) and (ii) respectively.  The circled cross-section in Fig. \ref{fig1} is magnified and shown for all the electric field contours. \color{black} The bias voltage is taken to be 1500 V. Ionization energy of Nitrogen in $\beta$-Ga$_2$O$_3$ is considered to be 2 eV \cite{tadjer2019}. The maximum electric field is at the metal edge in both the cases and we can see that the field is reduced in the guard ring structure in Fig. \ref{fig2}(b) as expected and also experimentally reported \cite{lin2019}.

\begin{figure}[t]
\centering
\includegraphics[width=3.1in, height=2.5in]{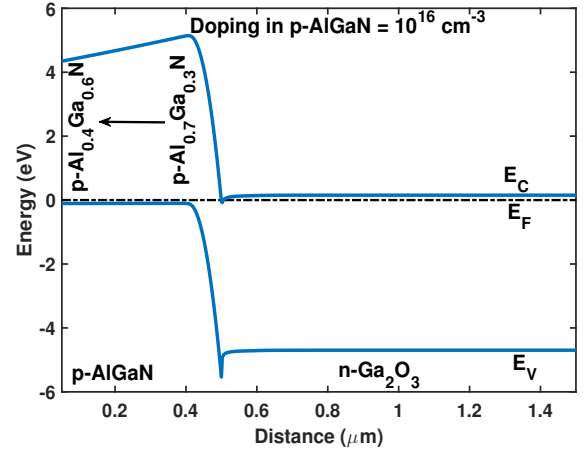}
\caption{\footnotesize Equilibrium energy band diagram of the simulated SBD with polarization doped p-AlGaN GR for T$_{GR}$ = 0.5 $\mu$m and N$_D$ = 10$^{16}$ cm$^{-3}$. Al composition is graded from 70 $\%$ to 40 $\%$ from the p-AlGaN/n-Ga$_2$O$_3$ heterointerface to the SBD surface.}
\label{fig4}
\vspace{-0.2cm}
\end{figure}

\begin{figure}[t]
\centering
\includegraphics[width=3.1in, height=2.5in]{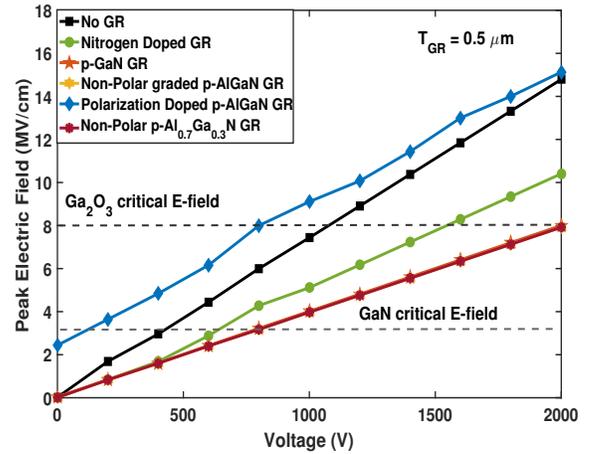}
\caption{\footnotesize Peak electric field vs applied bias in SBD for different GR configurations with 0.5 $\mu m$ of GR thickness and $10^{20}$ $cm^{-3}$ of GR doping.}
\label{fig5}
\vspace{-0.2cm}
\end{figure}

\begin{figure}[t]
\centering
\includegraphics[width=3in, height=2.3in]{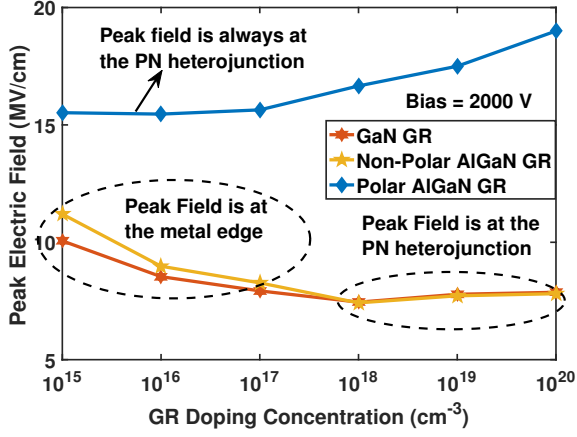}
\caption{\footnotesize Peak electric field vs GR doping concentration with 0.5 $\mu m$ of GR thickness and at applied bias of 2000 V for different GR configurations.}
\label{fig6}
\vspace{-0.2cm}
\end{figure}

We now explore the use of p-GaN as the guard ring material (Fig. \ref{fig3}(a)). The use a of magnesium-doped GaN guard ring, enables screening of the electric field at the metal edges due to the presence of mobile holes. The activation energy for Mg in case of p-GaN is considered to be 0.2 eV \cite{PhysRevB.97}. The doping concentration is taken to be $10^{20}$ $cm^{-3}$ (hole concentration = 8.2$\times$ 10$^{18}$ $cm^{-3}$) and the guard ring thickness is 0.5 $\mu m$ and the anode voltage is 2000 V. The peak electric field has now moved to the p-GaN/n-Ga$_2$O$_3$ heterojunction as can be seen in the electric field contour shown in Fig. \ref{fig3}(a). Fig. \ref{fig3}(b) shows the equilibrium energy band diagram of p-GaN/n-Ga$_2$O$_3$ heterojunction. At the anode bias of 2000 V, the peak electric field at the p-GaN/n-Ga2O3 heterojunction exceeds the critical field of GaN (3.3 MV/cm). We simulated the p-GaN guard ring configuration as a function of anode voltage and the results are presented in Fig \ref{fig5}. The  breakdown voltage of the p-GaN guard ring configuration presented here is 750 V.

To be able to leverage the high critical E-field of $\beta$-Ga$_2$O$_3$, a better design would be to use a guard ring material with an enhanced critical field as compared to $\beta$-Ga$_2$O$_3$ . So, we now study a graded p-AlGaN guard ring with aluminum composition graded from 70 $\%$ at the AlGaN/Ga$_2$O$_3$ interface to 40 $\%$ at the SBD surface. Lower Aluminum content is employed closer to the surface so as to achieve higher mobile hole concentration. Fig. \ref{fig3}(c) shows the electric field distribution using graded AlGaN for a doping concentration of $10^{20}$ $cm^{-3}$. However since the ionization energy of Mg is very high for AlGaN with high Al composition, it is very difficult to realize a high hole concentration. In fact for Mg concentration below $10^{18}$ $cm^{-3}$, the peak electric field is always at the metal edge because of the depletion of the entire GR. But if we grow c-axis oriented AlGaN with graded Al composition with decreasing Al composition from heterointerface to the surface, we can realize a 3-D slab of holes (3DHS) \cite{simon2010} due to the  polarization doping effect. This is expected to significantly increase the hole concentration even with low Mg doping due to field ionization of dopants. Fig. \ref{fig3}(e) shows the electric field distribution in the polarization doped p-Al$_x$Ga$_{(1-x)}$N (where x = 70 $\%$ at the heterojunction to 40 $\%$ at the SBD surface) GR. The hole concentration in this guard ring structure increased from the non-polar case by a significant amount (2$\times$ 10$^{17}$ to 6$\times$ 10$^{18}$ $cm^{-3}$). Band diagram of the polarization-doped polar graded p-AlGaN guard ring configuration is shown in Fig. \ref{fig3}(f). It can be observed that the peak electric field is now at the heterojunction mainly due to the positive polarization sheet charge at the p-Al$_{0.7}$Ga$_{0.3}$N/n-Ga$_2$O$_3$ heterojunction. The energy bands fall rapidly at the polar p-AlGaN/n-Ga$_2$O$_3$ interface, thus increasing the electric field compared to the non-polar p-AlGaN/n-Ga$_2$O$_3$ interface as shown in Fig. \ref{fig3}(g). Even for the low doped case, the bands fall rapidly at the p-AlGaN/n-Ga$_2$O$_3$ heterointerface as shown in Fig. \ref{fig4}. So, a polarization doped p-AlGaN GR would serve no benefit in reducing the peak electric field.

\begin{figure}[t]
\centering
\subfigure[]{
\includegraphics[width=1.71in, height=1.35in]{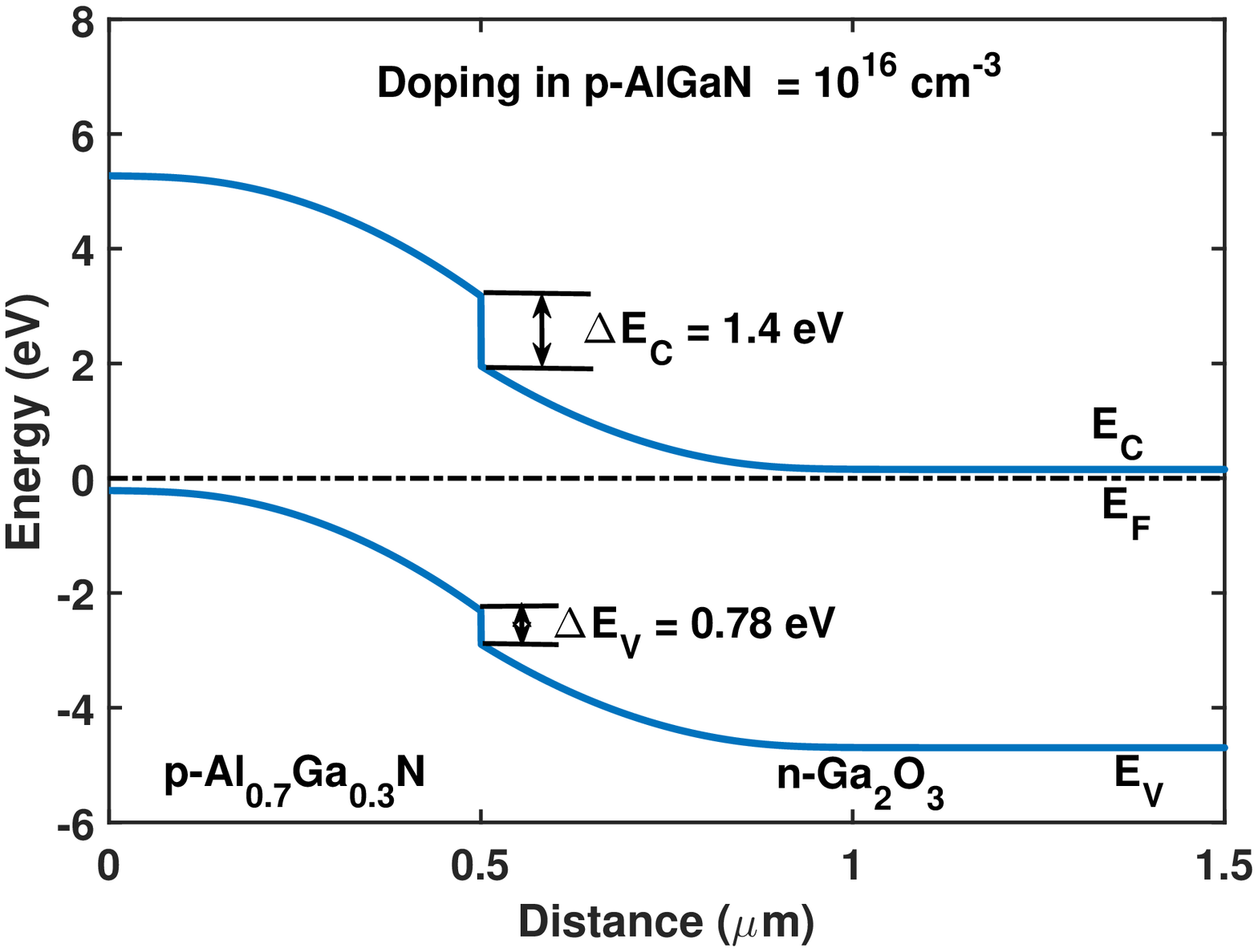}}\hspace{-0.2cm}
\subfigure[]{
\includegraphics[width=1.71in, height=1.35in]{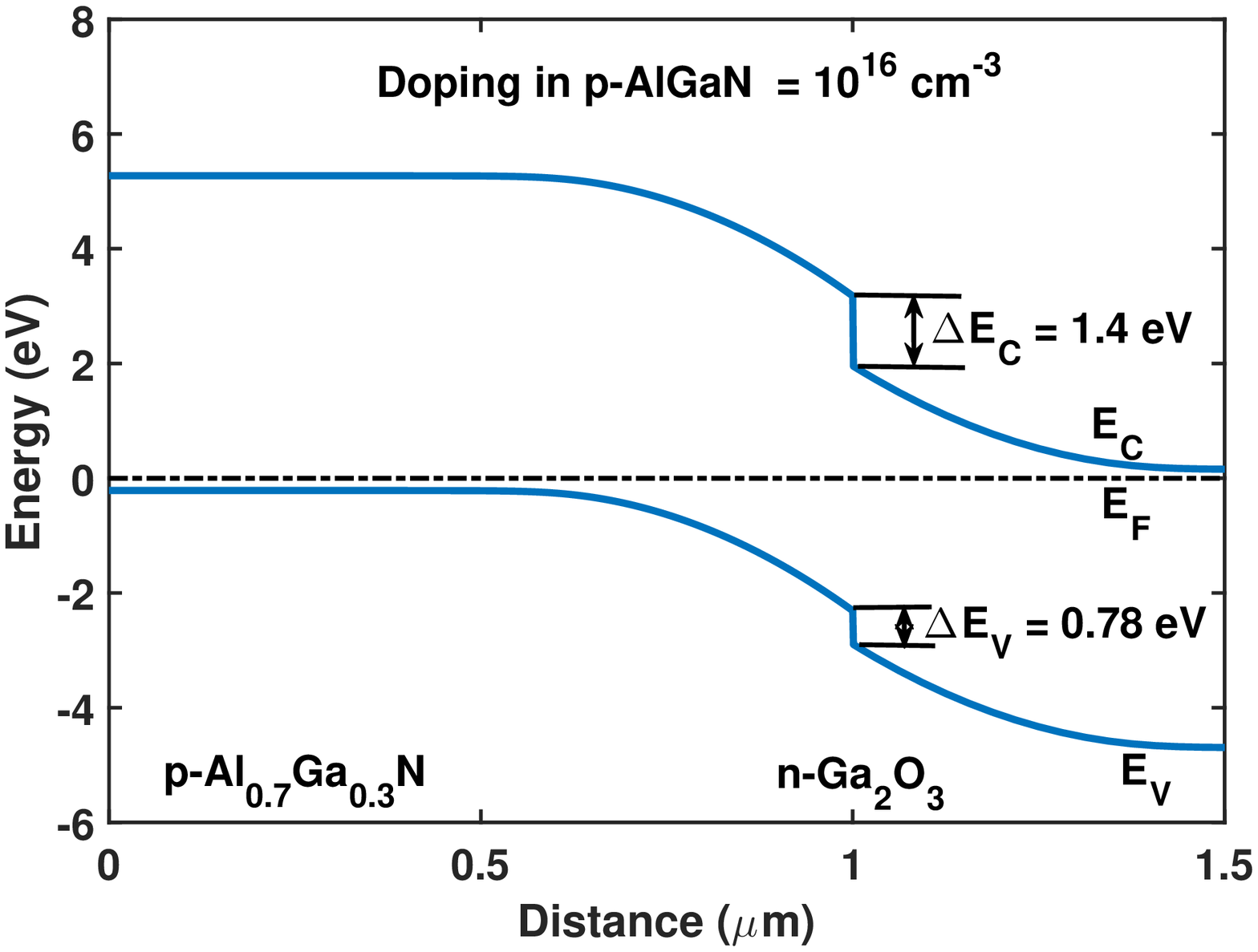}}
\\
\subfigure[]{
\includegraphics[width=3in, height=2.3in]{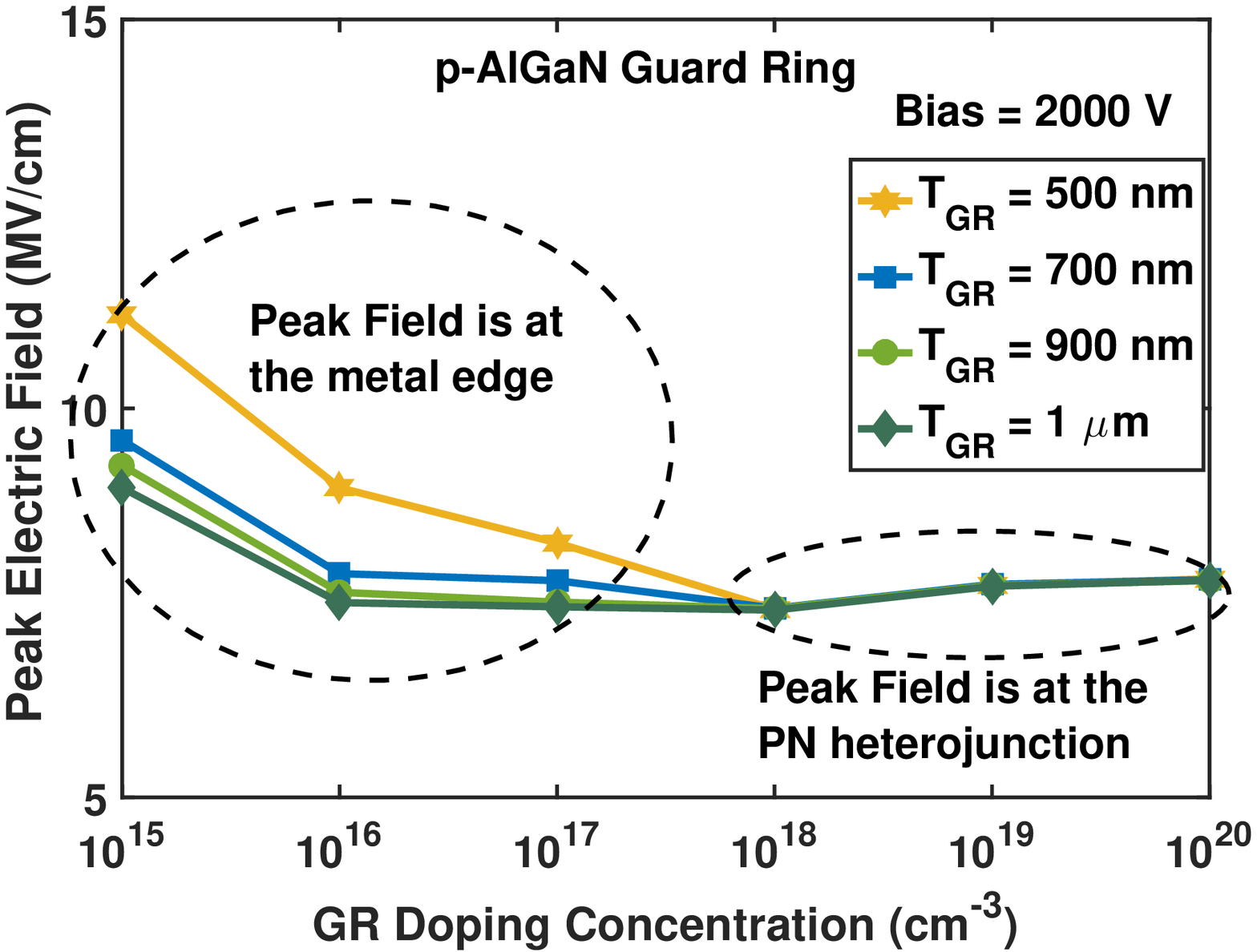}}
\caption{\footnotesize Equilibrium energy band diagram of the simulated SBD with nonpolar p-AlGaN GR for (a) T$_{GR}$ = 0.5 $\mu$m (b) T$_{GR}$ = 1 $\mu$m. Al composition is considered to be 70 $\%$. (c) Peak electric field vs GR doping concentration with non-polar AlGaN GR and at applied bias of 2000 V for different GR thicknesses.}
\label{fig7}
\vspace{-0.2cm}
\end{figure}

Fig. \ref{fig5} shows the peak electric field vs applied bias for SBD with the five different guard ring structures. The doping concentration for p-type guard ring is taken to be of $10^{20}$ $cm^{-3}$ and for Nitrogen-doped case it is taken to be of $10^{16}$ $cm^{-3}$ and the GR thickness is taken to be 0.5 $\mu$m.  The non-polar graded p-AlGaN, ungraded p-AlGaN, and p-GaN guard ring shows best performance in terms of reducing the peak electric field. However the SBD with GaN guard ring crosses it’s critical electric field of 3.3 MV at 750 V. The non polar p-Al$_x$Ga$_{1-x}$N GR with uniform Al composition (x = 60$\%$) has a critical electric field as high as gallium oxide and hence the breakdown voltage can be as high as 2000 V as shown in the Fig. \ref{fig5}. \color{black} The electric field is very high in case of the SBD with polarization doped AlGaN GR because of the high field at the heterointerface.

\begin{figure}[t]
\centering
\subfigure[]{
\includegraphics[width=3.5in, height=1.7in]{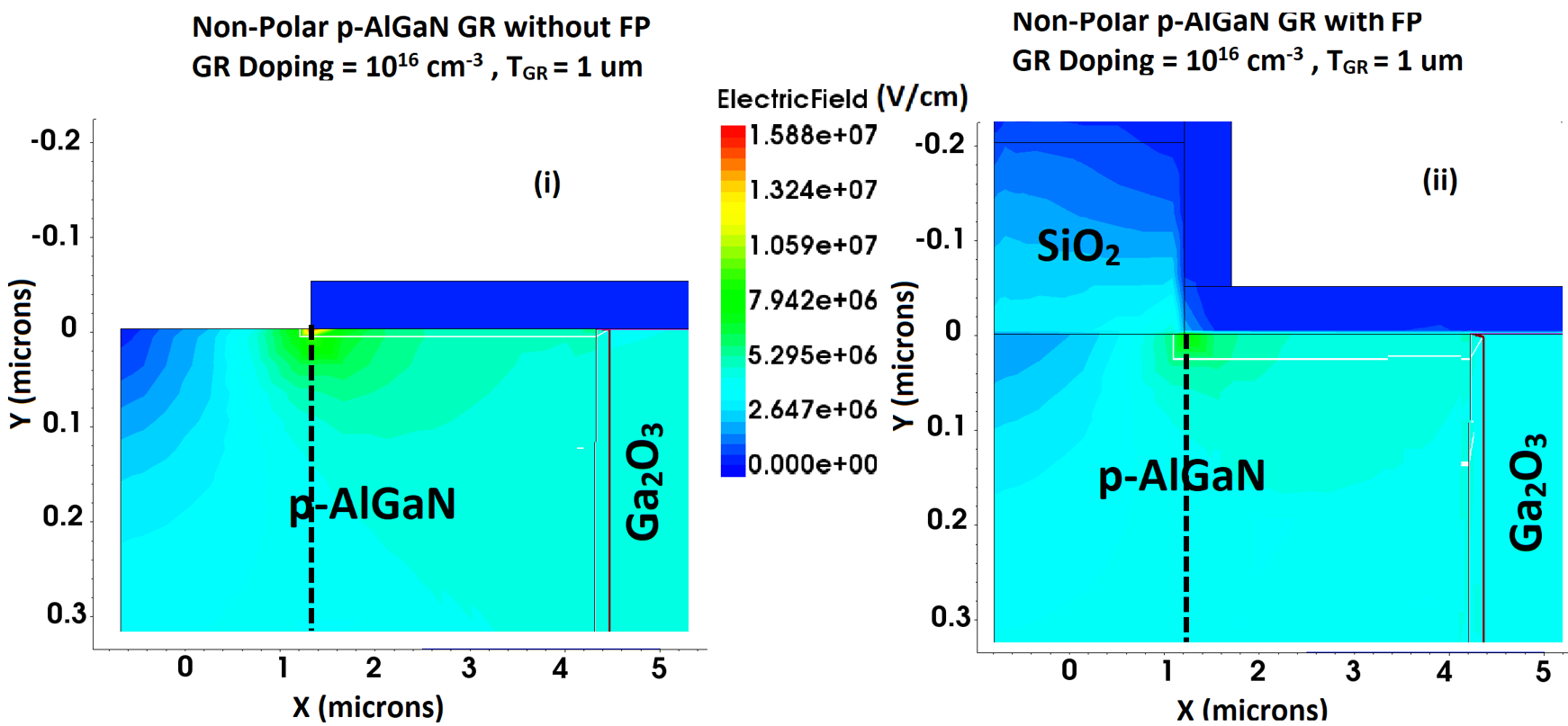}}\hspace{-0.2cm} 
\subfigure[]{
\includegraphics[width=3in, height=2.3in]{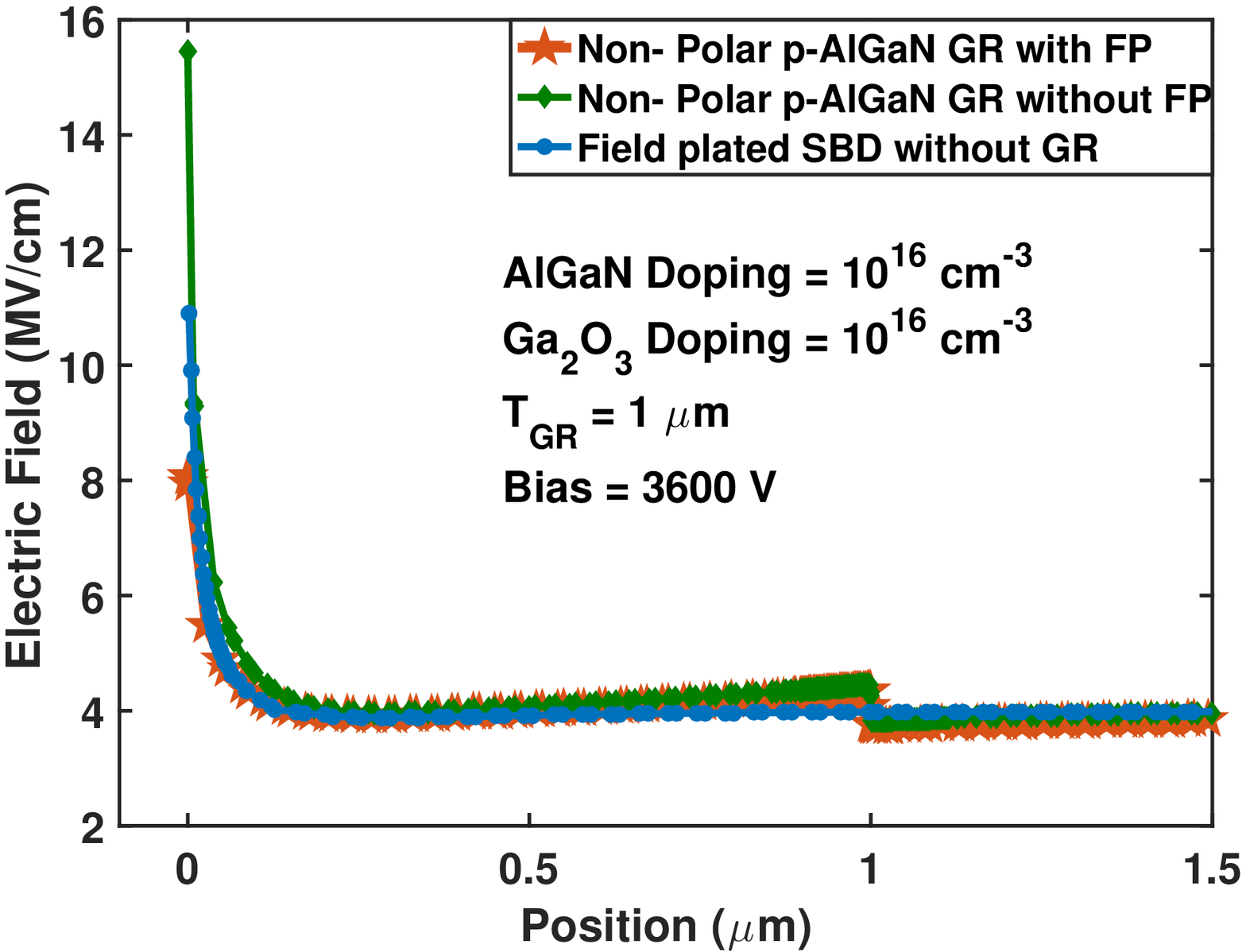}}\hspace{-0.2cm}
\caption{\footnotesize (a) Electric field distribution in SBD for 1 $\mu$m thick non-polar AlGaN GR  (i) without field plate (ii) with field plate at bias voltage of 3600 V and GR doping of $10^{16}$ $cm^{-3}$ (b) Electric field vs position for the two GR structure and also for field plated SBD with no GR at 3600 V and at GR doping of $10^{16}$ $cm^{-3}$ along the cutlines shown in the contour plots.}
\label{fig8}
\vspace{-0.2cm}
\end{figure}

Doping and thickness of the guard ring are critical parameters that would determine the amount of electric field screening and location of the peak electric field. We simulated the three guard ring configurations excluding the SBD with no GR and nitrogen-doped GR, as a function of Mg doping for a fixed thickness of 0.5 $\mu$m at a bias of 2000 V. Doping in the guard ring is found to determine the location of the peak electric field as shown in Fig. \ref{fig6}. For the polar p-AlGaN guard ring, the peak electric field is always at the heterojunction irrespective of the doping. In the case of p-GaN and non-polar p-AlGaN guard ring, the peak field is at the metal edge for doping concentrations lower than $10^{18}$ $cm^{-3}$.  Since a high GR doping is not necessary to minimize the peak field at the metal edge and since non polar graded and ungraded p-AlGaN GR shows similar performance, compositional grading in the GR is not required, which will mitigate the challenge involving growth of graded epitaxial layer of AlGaN inside the pocket of Ga$_2$O$_3$.

\color{black}
The electric field simulations clearly establish the comparably superior performance of non-polar p-AlGaN guard ring configuration. We now further focus on this particular configuration and study the effect of thickness of the guard ring. In the case of low-doped guard rings, in order to take advantage of holes, it is essential that the the thickness must be sufficiently large to realize an undepleted region close to the metal. Equilibrium energy band diagram of non-polar graded p-AlGaN GR with low doping ($10^{16}$ $cm^{-3}$) with two different thicknesses is shown in Fig. \ref{fig7}(a) and Fig. \ref{fig7}(b).  Presence of a quasi neutral region near the metal by employing a thick low doped guard ring (Fig. \ref{fig7}(b)) is expected to be beneficial for electric field screening. The quasi neutral region near the metal edge provides room for the growth of the depletion region at high reverse bias. \color{black} The effect of guard ring thickness as a function of Mg doping is summarized in Fig. \ref{fig7}(c). The peak electric field at a bias of 2000 V can be reduced from 9 MV/cm to 7.5 MV/cm at a guard ring doping of $10^{16}$ $cm^{-3}$. The peak electric field reduces as the doping increases for doping concentration less than $10^{18}$ $cm^{-3}$. As the GR thickness increases, the peak electric field reduces till a doping concentration of $10^{17}$ $cm^{-3}$. Above this concentration the peak electric field shifts to the pn-heterojunction and the GR thickness has no effect on the peak electric field. In this regime, there is no advantage of using a field plate, since the peak field region is buried.

\begin{figure}[t]
\centering
\includegraphics[width=3in, height=2.3in]{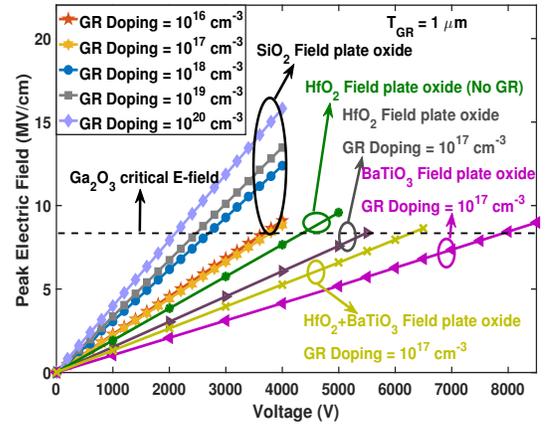}
\caption{\footnotesize Peak electric field vs applied bias in SBD with field plate and non-polar AlGaN GR for different GR doping concentrations and with field plate oxides with different dielectric constants. Also compared with standard field plated SBD structure with no GR.}
\label{fig10}
\vspace{-0.2cm}
\end{figure}


To further improve electric field management, we choose guard ring configuration B with a thickness of 1 $\mu$m and doping concentration of 10$^{16}$ cm$^{-3}$. In this case, the peak electric field is at the metal edge. Now we analyze the potential for further improvement in breakdown voltage with a field plate. Fig. \ref{fig8}(a) shows the electric field distribution in the SBD for a GR thickness of 1 $\mu$m and at a bias voltage of 3600 V (i) without  and (ii) with a field plate respectively. An SiO$_2$ layer of 200 nm is used as the field plate oxide.  We also compared our design with a field plated SBD without GR as shown in \ref{fig8}(b). The field plating has more impact on the reduction of the peak electric field compared to the SBD with only GR. However, SBD having a field plate combined with a thick p-AlGaN GR dramatically reduces the peak electric field at the metal edge. \color{black} Here again the peak electric field reduces as the doping increases till doping concentration reaches $10^{17}$ $cm^{-3}$. So, the best design would be a thick low doped p-AlGaN GR with a field plate. 

The Fig. \ref{fig10} shows the effect of field plating on reducing the peak electric field for the SBD with non-polar p-AlGaN GR.  In Fig. \ref{fig10}, we can see that for doping concentration less than $10^{18}$ $cm^{-3}$, in the case of non-polar p-AlGaN GR, field plating significantly reduces the electric field at the metal edge and the electric field crosses the $\beta$-Ga$_2$O$_3$ critical electric field of 8 MV/cm at a bias voltage of around 3600 V. For doping concentration above $10^{17}$ $cm^{-3}$, the field plating has no effect on reducing the electric field because of shifting of the peak electric field from metal edge to the pn-heterojunction.  We can also see that use of a high-k dielectric (HfO$_2$, in this case with relative permittivity of 22) with same dimension as the previous case as field plate oxide in conjunction with the guard ring reduces the peak electric field dramatically and the device reaches the $\beta$-Ga$_2$O$_3$ critical electric field of 8 MV/cm at a reverse bias voltage of 5200 V, which is significantly higher than the highest reported breakdown voltage for any vertical SBD with 10 $\mu m$ drift layer \cite{huili}. The high permittivity difference between the semiconductor and the dielectric generates polarization bound charge inside the dielectric which balances the depletion charge at the semiconductor interface \cite{xia2019design,rajan2020dielectric,kabemura2018enhancement}.  This charge balance results in flattening of the electric field profile at the dielectric/semiconductor interface reducing its peak magnitude. We have also simulated a field plated SBD with HfO$_2$ as field plated oxide without a GR and it achieves a breakdown voltage of 4300 V. So, the use of a GR in conjunction with a field plate increases the breakdown voltage by 900 V compared to the field plated SBD with no GR. We have also analyzed other extreme high-k material such as BaTiO$_3$ (relative permittivity of 300) as field plate oxide and the breakdown voltage was found to reach 7800 V when used in conjunction with GR. Since BaTiO$_3$ has no conduction band offset with $\beta$-Ga$_2$O$_3$, use of BaTiO$_3$ underneath the metal might cause charge trapping at the dielectric/semiconductor interface. To mitigate the charge trapping, we have analyzed a stacked dielectric of HfO$_2$ (5 nm)+BaTiO$_3$ (295 nm). Here the large thickness ratio is used to maintain the high dielectric constant for the series configuration which results in an effective dielectric constant of 248. Using this structure in conjunction with the thick guard ring we are able to get a breakdown voltage of 6200 V.

\color{black}
Among all the devices described the SBD with non-polar p-AlGaN GR added with field plate and a high-k field plate oxide is found out to be best choice to reduce the electric field.  The thick $\beta$-Ga$_2$O$_3$ epilayers (10$\mu$m) on (010) $\beta$-Ga$_2$O$_3$ substrates used in this design can be grown using HVPE and has already been demonstrated \cite{murakami2014homoepitaxial,leach2019halide}. Selective area epitaxy of AlGaN GRs inside Ga$_2$O$_3$ trench pockets can be done using MOCVD \cite{kapolnek1997anisotropic,kawaguchi1998selective}. We understand AlGaN heteroepitaxy on Ga$_2$O$_3$ could lead to compromised material quality and will require extensive growth and process optimizations. Experimentally, trench Ga$_2$O$_3$ SBDs with field plate structures, were able to achieve FOM as high as 0.95 GW/cm$^2$ \cite{huili}. Using the concepts explored in this work, if Ga$_2$O$_3$ SBDs with GR in conjunction with field plates are implemented, we expect this design to surpass the already high FOM achieved with Ga$_2$O$_3$ power SBDs. For instance, with a breakdown voltage of 6200 V and an estimated R$_{ON,SP}$ of 3.55 m$\Omega$-cm$^2$, assuming a mobility of 176 cm$^2$/V.s \cite{zhang2019}, we should be able to achieve extremely high FOM of 10.8 GW/cm$^2$.

\color{black}
\vspace{-0.2cm}
\section{Conclusion} 
\label{sec4} 
A novel approach to reduce the electric field and thus increasing the breakdown voltage for schottky barrier diode by using p-doped III-nitride guard ring is proposed and shown through a detailed device simulation. This approach circumvents the issue of lack of p-type doping in gallium oxide. The SBD with thick low doped non-polar p-AlGaN GR in conjunction with a field plate and a high-k dielectric can serve best in terms of reducing the peak electric field. The inclusion of field plate and a high permittivity field plate oxide in case of low doped GR is shown to further reduce the electric field at the metal edges. Further research into the interface properties of the AlGaN/Ga$_2$O$_3$ heterointerface will lead to better understanding and use of such heterojunction-based structures for high performance power electronic devices.

\section*{Acknowledgement}
This material is based upon work supported by the Air Force Office of Scientific Research under award number FA9550-18-1-0507 (Program Manager: Dr. Ali Sayir). Any opinions, finding, and conclusions or recommendations expressed in this material are those of the author(s) and do not necessarily reflect the views of the United States Air Force.

\bibliographystyle{IEEEtran}
\bibliography{ref}
\vspace{-0.5cm}
\end{document}